\documentclass[conference,subeqn,graphicx,psfig,amssymb,amsthm,amsmath]{IEEEtran}
\IEEEoverridecommandlockouts
\usepackage{epsfig,graphicx}
\usepackage{graphicx}
\usepackage{epsfig}
\usepackage{latexsym}
\usepackage{amsfonts}
\usepackage{rawfonts}
\usepackage[latin1]{inputenc}
\usepackage[T1]{fontenc}
\usepackage{calc}
\usepackage{url}
\usepackage{enumerate}
\usepackage{color}
\usepackage[tbtags]{amsmath}
\usepackage{amssymb}
\usepackage{upref}
\usepackage{times}
\usepackage{amsmath}
\usepackage{cite}
\usepackage{tabularx}
\newcommand{\lb} {\left}
\newcommand{\rb} {\right}
\newcommand{\nn} {\nonumber}
\begin{document}
\title{Relay Selection to Improve Secrecy in Cooperative Threshold Decode-and-Forward Relaying}

\author{ Chinmoy Kundu, Telex M. N. Ngatched, and Octavia A. Dobre \\Faculty of Engineering and Applied Science
\\Memorial University, St. John's, NL A1B 3X5, Canada\\ ckundu@mun.ca, 
tngatched@grenfell.mun.ca, and odobre@mun.ca}

% \author{Chinmoy Kundu}
% \affil{Bharti School of Telecomm. Technology \& Mgmt.\\Indian Institute of Technology, Delhi\\New Delhi-110016, India\\E-mail: chinmoy.kundu@dbst.iitd.ac.in}
% \author{Ranjan Bose}
% \affil{Dept. of Electrical Engineering\\Indian Institute of Technology, Delhi\\New Delhi-110016, India\\E-mail: rbose@ee.iitd.ac.in}

% \author[1]{Chinmoy Kundu}
% \author[2]{Ranjan Bose}
% \affil[1]{Bharti School of Telecomm. Technology \& Mgmt.\\Indian Institute of Technology, Delhi\\New Delhi-110016, India\\E-mail: chinmoy.kundu@dbst.iitd.ac.in}
% \affil[2]{Dept. of Electrical Engineering\\Indian Institute of Technology, Delhi\\New Delhi-110016, India\\E-mail: rbose@ee.iitd.ac.in}

% \thispagestyle{empty}
% \pagestyle{empty}

\maketitle
\begin{abstract}
In this paper, relay selection is considered to enhance security of a cooperative system with
multiple threshold-selection decode-and-forward (DF) relays. Threshold-selection DF relays
are the relays in which a predefined signal-to-noise ratio is set for the condition of
successful decoding of the source message. We focus on the practical and general scenario
where the channels suffer from independent non-identical Rayleigh fading and where the direct
links between the source and destination and source and eavesdropper are available. Based on
channel state information knowledge, three relay selection strategies, namely traditional, improved traditional, and optimal, 
are studied. In particular, the secrecy outage probability of all three strategies are obtained in closed-form. It is found
that the diversity of secrecy outage probability of all strategies can improve with increasing the
number of relays. It is also observed that the secrecy outage probability is limited by either 
the source to relay or relay to destination channel quality.
\end{abstract}

\section{Introduction}

To secure the broadcast nature of wireless communication against eavesdroppers, physical
layer security has gained much prominence \cite{wyner_wiretap}. Motivated by
recent advances in cooperative communication systems \cite{Laneman_Wornell_cooperative_diversity}, employing 
the cooperative technique
to enhance physical layer security of wireless systems has recently been receiving significant
research interest. Compared to multi-relay assisted transmission, relay selection in which a
single relay among all possible candidates is selected for relaying a source's signal has been
shown to optimize system resource utilization, such as power and bandwidth, while maintaining
the same diversity order.

Relay selection to improve secrecy in cooperative communication system has received considerable
attention recently \cite{krikidis_iet_opport_rel_sel, 
krikidis_twc_Rel_Sel_Jam, Zou_Wang_Shen_optimal_relay_sel, 
Alotaibi_Relay_Selection_MultiDestination, Bao_Relay_Selection_Schemes_Dual_Hop_Security, Poor_Security_Enhancement_Cooperative, 
Fan_Karagiannidis_Secure_Multiuser_Communications, 
jindal_kundu_secrecy_paper_vtc_2014, jindal_kundu_secrecy_af_coml, Kundu_relsel, Qahtani_relsel}. The relays considered in these works are conventional
amplify-and-forward (AF) or decode-and-forward (DF) relays \cite{Laneman_Wornell_cooperative_diversity}. In all these works,
relay selection is performed depending on the availability of the instantaneous channel state information (ICSI) or 
statistical channel state information (SCSI) of the links. Based on the knowledge
of ICSI or/and SCSI, the following three cases have been considered mostly for the relay selection
problem. Case i): when the ICSI of the source to relay and relay to destination is known. In this
case, the selected relay is the one which achieves the maximum rate through the source-relay-destination channel, which is 
described as the main channel in physical layer security. We refer to this as traditional relay selection (TS). Case ii): 
in addition to the ICSI of source to relay and relay to destination, 
the SCSI of the relay to eavesdropper channels are known. This is an improvement over the previous case
and we refer to it as improved traditional selection (ITS). Case iii): when the ICSI of all links
are known. In this case, the relay selected is the one which provides maximum secrecy rate. We refer to this as optimal selection (OS). 
In practise, the ICSI of the various links can be acquired
using one of the techniques described in \cite{Khisti_a_simple_cooperative_diversity} and the references therein.

With the notable exception of \cite{Kundu_relsel},  \cite{Qahtani_relsel} and \cite{sarbani_Kundu_threshold_relay}, most of the existing work on physical
layer security in DF relay cooperative systems has only considered the high signal-to-noise ratio
(SNR) regime for the source to relay link. This is not very practical as fading can severely
degrade the channel quality of a link in wireless communication systems. In \cite{Kundu_relsel}, the source
to relay link quality is taken into account by considering that the rate at the destination is
limited by the minimum of the source to relay and relay to destination rate. In \cite{Qahtani_relsel}, the set of
successful relays which recover the source symbol are those for which the source to relay link
rate is above a minimum threshold rate. Furthermore, only \cite{Qahtani_relsel} has considered the existence
of direct links from source to destination and source to eavesdropper. However, in this work, all the
links are assumed to experience independent and identical Rayleigh fading. Though the identical
distribution assumption makes the analysis more tractable, it may not be valid for practical
wireless communication applications because, in general, the relays are not closely placed in
real environments. Moreover, only the TS scheme is studied in \cite{Qahtani_relsel} and the relay selection problem is 
not tackled in \cite{sarbani_Kundu_threshold_relay}.

With this motivation in mind, in this paper, we study relay selection to enhance security of a
cooperative system with multiple threshold-selection DF relays. In threshold-selection relaying
scheme for DF cooperation protocol, the possible candidate relays for selection are those for
which the signal-to-noise ratio (SNR) is above a predefined threshold \cite{liu_opt_threshold}. We consider the
more practical scenario where the direct links between the source to destination and source to
eavesdropper exist and where the links experience independent but not necessarily identically
distributed Rayleigh fading. The main contributions of the paper can be summarized as follows:

\begin{enumerate}
 \item We study three relay selection strategies, depending on the ICSI and SCSI knowledge to
 enhance the secrecy outage probability of threshold-selection relaying.
  \item We derive the secrecy outage probability in closed-form for the most general case of
independent but non-identical channels.
\item We obtain the secrecy outage assuming direct links from source to destination and source
to eavesdropper.
\end{enumerate}

In detailing our contributions, we observe that the studied relay selection strategies can 
increase the diversity order of secrecy outage probability with increasing the number of relays.
Interestingly, we also observe that the secrecy outage probability can not be increased beyond a
certain level if either the source to relay or relay to destination channel quality is kept fixed
while the other is increased.

The rest of the paper is organized as follows. Section \ref{sec_system} introduces the system model. 
Section \ref{sec_multrel} evaluates 
the secrecy outage probability for the relay selection strategies. Section \ref{sec_results} 
discusses the results, and finally, Section \ref{sec_conclusions} 
concludes the paper.

\textit{Notation:} $\mathbb{P}[\cdot]$ is the probability of occurrence 
of an event, $\mathbb{E}_X[\cdot]$ defines the expectation of its argument over the random variable (r.v) 
$X$, $(x)^+\triangleq \max(0,x)$ and $\max{(\cdot)}$ denotes the maximum 
of its argument, $F_X (\cdot)$ represents the 
cumulative distribution function (CDF) of the r.v $X$, and 
$f_X (\cdot)$ is the corresponding probability density function (PDF).

\section{System Model}
\label{sec_system}
\begin{figure}
\centering
\includegraphics[width=0.18\textwidth] {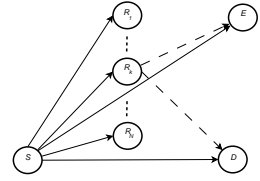}
\vspace*{-0mm}
\caption{System model with multiple threshold-selection DF relays. }
\label{FIG_1}
\vspace*{-6mm}
\end{figure}

The system model consists of one source ($S$), one destination ($D$), 
one passive eavesdropper ($E$), and $N$ DF relays ($R_k$, $k\in \{1, 2, \cdots N\} $), as shown in Fig. \ref{FIG_1}. 
All 	nodes are equipped with a single antenna. The relays are half-duplex in nature, and hence, complete 
information transmission takes place in two time slots. 
$S$ broadcasts its message in the first time slot. We assume that the relays are threshold-selection DF 
type \cite{liu_opt_threshold}; in other words, they 
correctly decode the received message and retransmits in the second time slot only if their SNR is above a 
threshold, $\gamma_{th}$. 
% It does not transmit received message at all if it cannot decode the message correctly. 
The SNR threshold, $\gamma_{th}$, can be properly chosen to 
achieve the goal of correct decoding. The channels are modeled as 
independent non-identically distributed flat Rayleigh fading. 
Both $D$ and $E$ utilize maximal ratio combining (MRC) technique 
to get the advantage of two copies of same signal from the direct transmission and the relayed transmission. 
The received SNR, $\gamma_{xy}$, of any arbitrary 
$x$-$y$ link from node $x$ to node $y$ can be expressed as
% \begin{align}
% \label{eq_1} 
$\gamma_{xy} = \frac{P_{x}|h_{xy}|^2 }{N_{0_{y}}}$,
% \end{align}
where $x$ and $y$ are from the set $\{S, R, D, E\}$ for any possible combination of $x$-$y$, $x\ne y$. $P_{x}$ is 
the transmit power from node $x$ and $N_{0_{y}}$ is the noise variance of the 
additive white Gaussian noise (AWGN) at node $y$. As $h_{xy}$ is assumed Rayleigh distributed with 
average power unity, i.e., $\mathbb{E}[|h_{xy}|^2]=1$, $\gamma_{xy}$ is exponentially distributed 
with mean $1/\lambda_{xy}=P_{x}/N_{0_{y}}$. The CDF
of $\gamma_{xy}$ can be written as 
% \begin{align}
% \label{eq_2} 
$F_{\gamma_{xy}}(z)=1-\exp(-\lambda_{xy}z), z\geq 0$. 
% \end{align}

For simplicity, we further denote the parameters of the $S$-$E$ and $R_k$-$E$ links which are terminating at $E$
as $\lambda_{SE}=\alpha_{se}$ and $\lambda_{R_kE}=\alpha_{ke}$, respectively. 
Parameters of the other links which are conveying messages towards $D$, i.e., $S$-$R_k$ or $R_k$-$D$, 
are denoted by $\lambda_{SR_k}=\beta_{sk}$ and $\lambda_{R_kD}=\beta_{kd}$, respectively. 

The achievable secrecy rate of the system is \cite{wyner_wiretap} 
\begin{align}
\label{eq_3}
C_{S}\triangleq{\frac{1}{2}\lb[\log_2\lb(\frac{1+\gamma_{M}}{1+\gamma_{E}}\rb)\rb]}^+,
\end{align} 
where $\gamma_{M}$ and  $\gamma_{E}$ are the main channel and eavesdropper channel SNR at $D$ and $E$, respectively. 
The term $1/2$ reflects the fact that two time slots are required for information transfer. 

The secrecy outage probability of the system represents the probability that the achievable secrecy rate 
is less than a target secrecy rate, $R_s$, and is expressed as
\begin{align} 
\label{eq_4}
P_o\lb(R_s\rb)&=\mathbb{P}\left[C_{S} < R_s  \right]=
\mathbb{P}\lb[\gamma_{M} < \rho\lb(1+\gamma_{E}\rb)-1 \rb], 
% &=\mathbb{E}_{\gamma_E}\lb[F_{\gamma_M}\lb(\rho\lb(1+\gamma_E\rb)-1\rb)\rb]
\end{align} 
where $\rho = 2^{2R_s}$.

% Depending on ICSI and SCSI knowledge of the links, three different relay selection schemes are proposed: 
% i) TS, ii) ITS, and iii) OS, and corresponding secrecy outage probabilities are evaluated. 
% ICSI of $R_k$-$E$ links is not required for the TS and ITS schemes except the OS scheme for all $k$.
% However, SCSI of $R_k$-$E$ links is required only for the ITS scheme. 

\section{Secrecy Outage of Relay Selection}
\label{sec_multrel}
Let us assume that $\mathbb{S}$ is a set representing the relays which are able to decode successfully at the first stage. 
A relay is to be selected from this particular set by a relay selection rule in the second stage. 
% To find out the secrecy outage probability of the system, we need to first find out all possible combinations of relays in the set including empty set and 
% corresponding secrecy outage probabilities conditioned on the set $\mathbb{S}$.
% Due to the law of total probability we need to then sum over all such probabilities starting from no relays in the set to the $N$ relays in the set. 
The secrecy outage probability of the system with relay selection, $P_o(R_s)$, can be mathematically represented as 
\begin{align}
\label{eq_58_A}
P_o(R_s)=\sum\limits_{K=0}^N   \sum   \mathbb{P} [\mathbb{S}] P_o^{\mathbb{S}}(R_s),
\end{align}
where $\mathbb{P} [\mathbb{S}]$ represents the probability of occurrence of a particular set $\mathbb{S}$ containing $K$ relays, 
and $P_o^{\mathbb{S}}(R_s)$ represents the secrecy outage probability for a given relay selection rule.
The second summation in  (\ref{eq_58_A}) must  be performed for $ N \choose K$ possible combinations. 
It should be noted that $\mathbb{P} [\mathbb{S}]$ and $P_o^{\mathbb{S}}(R_s)$ can be evaluated independently, 
as they are the result of 
independent events.

Now let us assume that the relays which are unable to exceed $\gamma_{th}$ constitute the set $\bar{\mathbb{S}}$.
$\mathbb{P} [\mathbb{S}]$ can be easily evaluated by multiplying the probability 
of occurrence of $\mathbb{S}$ and $\bar{\mathbb{S}}$. 
With the probability that a particular relay $k$ is in 
$\mathbb{S}$ given by   
\begin{align}
\label{eq_th1}
\mathbb{P}[\gamma_{sk}>\gamma_{th}]=\exp(-\beta_{kd}\gamma_{th}), 
\end{align}
$\mathbb{P}[\mathbb{S}]$ can be evaluated as
\begin{align}
\label{eq_th2}
\mathbb{P}[\mathbb{S}] = \prod_{\forall k \in \mathbb{S}}^K \exp(-\beta_{sk}\gamma_{th})
\prod_{\forall j \in \bar{\mathbb{S}}}^{N-K} (1-\exp(-\beta_{sj}\gamma_{th})).
\end{align}

When $\mathbb{S}$ is the empty set, there exists only the direct link to $D$ and $E$ from $S$. 
In this case, the secrecy outage probability can be obtained from (\ref{eq_4}) as
\begin{align}
\label{eq_nosel} 
P_o^{\mathbb{S}}(R_s) = 1 - \frac{\alpha_{se}\exp{\lb(-\beta_{sd}(\rho-1)\rb)}}{\rho \beta_{sd}+\alpha_{se}}.
\end{align}
When $\mathbb{S}$ contains a single relay, the secrecy outage probability is obtained 
by using $\gamma_M=\gamma_{sd}+\gamma_{kd}$ and $\gamma_E=\gamma_{se}+\gamma_{ke}$.
The $\gamma_M$ and $\gamma_E$ distributions can be readily found for different parameters of the exponential r.vs as 
\cite{sum_expo_mohamed_akkouchi}
\begin{align}
\label{eq_sumexpo}
f_X(x)=B_1 e^{-\lambda_1 x}
+B_2  e^{-\lambda_2 x}, 
\end{align}
where 
% \begin{align}
% \label{eq_b1b2}
$B_1=\frac{\lambda_2\lambda_1}{\lambda_2-\lambda_1}$, $B_2=\frac{\lambda_2\lambda_1}{\lambda_1-\lambda_2}$.
% \end{align}
$\lambda_1$ and $\lambda_2$ are the parameters of the two independent exponentially distributed r.vs, 
with $\lambda_1\ne\lambda_2$.
% \begin{align}
% \label{eq_b1b2}
% B_1=\frac{\alpha_{ke}\alpha_{se}}{\alpha_{ke}-\alpha_{se}};~~~ 
% B_2=\frac{\alpha_{ke}\alpha_{se}}{\alpha_{se}-\alpha_{ke}}.
% \end{align}
Thus, the secrecy outage probability can be obtained from (\ref{eq_4}) as
\begin{align}
\label{eq_singlerel} 
P_o^{\mathbb{S}}(R_s) = 1 &- \frac{\beta_{sd}\alpha_{ke}\alpha_{se}\exp{\lb(-\beta_{kd}(\rho-1)\rb)}}{(\beta_{sd}-\beta_{rd})(\rho\beta_{kd}+\alpha_{re})(\rho\beta_{kd}+\alpha_{se})} \nn\\
&-\frac{\beta_{kd}\alpha_{ke}\alpha_{se}\exp{\lb(-\beta_{sd}(\rho-1)\rb)}}{(\beta_{rd}-\beta_{sd})(\rho\beta_{sd}+\alpha_{re})(\rho\beta_{sd}+\alpha_{se})}.
\end{align}

\subsection{Traditional Selection (TS)}
\label{sec_trad_se_sd}
The traditional relay selection rule does not take into account the $R_k$-$E$ channel quality for all k. This 
scheme selects the relay which achieves the highest rate through the $R_k$-$D$ link, as successful decoding has 
already been performed in the first stage. The highest rate is achievable on the link having highest instantaneous SNR. 

The secrecy outage probability of the traditional rule corresponding to $\mathbb{S}$ can be 
evaluated using the law of total probability as 
\begin{align}
\label{eq_58}
&P_o^{\mathbb{S}}(R_s)=\sum_{\forall k \in \mathbb{S}}^K \mathbb{P}\lb[\text{Relay}=R_k\rb]\mathbb{P}\lb[C_S^k<R_s\rb]  \nn\\
&=\sum_{\forall k \in \mathbb{S}}^K\mathbb{P}\lb[\gamma_{kd}>\gamma_{kd}^{-}\rb]
\mathbb{P}\lb[\frac{1+\gamma_{kd}+\gamma_{sd}}{1+\gamma_{ke}+\gamma_{se}}\leq \rho\rb] \nn \\
&=\sum_{\forall k \in \mathbb{S}}^K \mathbb{P}\lb[\gamma_{kd}^{-}<\gamma_{kd}\leq \lb(\rho-1\rb)+\rho\lb(\gamma_{ke}+\gamma_{se}\rb)-\gamma_{sd}\rb],  
\end{align}
where $C_S^k$ is the secrecy outage probability for the $k$th relay in $\mathbb{S}$, $\gamma_{kd}^{-}= \max  \{ \gamma_{id}\}$, where $\forall i \in \mathbb{S}$ and $i\ne k$, is the 
maximum SNR between all relays which are not selected from $\mathbb{S}$ by the relay selection rule. The 
distribution $f_{\gamma_{kd}^{-}}(y)$ is expressed as \cite{Kundu_relsel}
\begin{align}
\label{eq_pdf}
 f_{\gamma_{kd}^-}(y)=-\sum\limits_{m=1}^{K-1}\lb(-1\rb)^{m}\mathbb {\sum}_m^{\prime} \beta_m^{\prime}e^{-y\beta_m^{\prime}},
\end{align} 
where ${\sum}_m^{\prime} $ is defined as
\begin{align}
\mathbb {\sum}_m^{\prime}=\sum_{\substack{i_1=1\\i_1\ne k}}^{K-(m-1)}
\sum_{\substack{i_2=i_1+1\\i_2\ne k}}^{K-(m-2)}\cdots \sum_{\substack{i_{m-1}=i_{m-2}+1\\i_{m-1}\ne k}}^{K-1} 
\sum_{\substack{i_m=i_{m-1}+1\\i_m\ne k}}^K,
\end{align}
and $\beta_m^{\prime}=\sum_{l=1}^m\beta_{{i_l}d}$. To further evaluate the probability in (\ref{eq_58}), let us assume a new r.v, $X=\gamma_{ke}+\gamma_{se}$. 
As all r.vs in (\ref{eq_58}) are independent, the solution can be written in integral form as 
\begin{align}
\label{eq_59_A}
P_o^{\mathbb{S}}(R_s)=\sum_{\forall k \in \mathbb{S}}^K
&\lb( I_1+I_2\rb),
\end{align}
where 
\begin{align}
\label{eq_i1_i2}
I_1&=\int_{(\rho-1)}^\infty \int_{\frac{z-(\rho-1)}{\rho}}^\infty \int_0^\lambda \int_y^\lambda 
f_{\gamma_{kd}}(t)f_{\gamma_{kd}^{-}}(y)f_{X}(x)\nn\\ &\times f_{\gamma_{sd}}(z) dt dy dx dz,\\
\label{eq_i1_i3}
I_2&=\int_{0}^{(\rho-1) }\int_{0}^\infty \int_0^\lambda \int_y^\lambda 
f_{\gamma_{kd}}(t)f_{\gamma_{kd}^{-}}(y)f_{X}(x)\nn\\
&\times f_{\gamma_{sd}}(z) dt dy dx dz,
\end{align}
and $\lambda=\rho-1+\rho x-z$. 

In deriving (\ref{eq_59_A}), $x$, $y$, and $z$ represent the realizations of $X$, $\gamma_{kd}^{-}$, and $\gamma_{sd}$, respectively. 
The distribution $f_{\gamma_{kd}^{-}}(y)$ is given in (\ref{eq_pdf}), while the distribution of the r.v $X$ is given in (\ref{eq_sumexpo}).  
The integration limits are due to following reasons: i) $\gamma_{kd}^{-}$ should be always less than $\lambda$; hence, $y$ takes values 
from zero to $\lambda$, ii) none of the r.vs can take negative values, and hence, when $\gamma_{sd}$ exceeds $(\rho-1)$, $X$ 
is higher than $(z-(\rho-1))/\rho$ in (\ref{eq_i1_i2}), iii) when  $\gamma_{sd}$ is below $(\rho-1)$, $X$ has positive values; hence, 
the corresponding integral is from zero to infinity in (\ref{eq_i1_i3}).
Final expressions of $I_1$ and  $I_2$ are given in (\ref{eq_td_i1}) and (\ref{eq_td_i2}), respectively.
\subsection{Improved Traditional Selection (ITS)}
\label{sec_improved_trad}
Traditional relay selection can be improved by using the statistical channel knowledge of the $R_k$-$E$ link, $\alpha_{ke}$, in (\ref{eq_58}) to obtain 
$P_o^{\mathbb{S}}(R_s)$, as follows
\begin{align}
\label{eq_tradimp} 
&P_o^{\mathbb{S}}(R_s)\nn\\
&=\sum_{\forall k \in \mathbb{S}}^K  \mathbb{P}\lb[\frac{\gamma_{kd}}{1/\alpha_{ke}}>\lb(\frac{\gamma_{kd}}{1/\alpha_{ke}}\rb)^{-}\rb]
\mathbb{P}\lb[\frac{1+\gamma_{kd}+\gamma_{sd}}{1+\gamma_{ke}+\gamma_{se}}\leq \rho\rb] \nn\\
&=\sum_{\forall k \in \mathbb{S}}^K  \mathbb{P}\lb[ \gamma_{kd}> \frac{\gamma_M ^-}{\alpha_{ke}}\rb]\mathbb{P}\lb[\gamma_{kd}< (\rho-1)+\rho\lb(\gamma_{ke}+\gamma_{se}\rb)-\gamma_{sd} \rb]\nn\\
&=\sum_{\forall k \in \mathbb{S}}^K (I_3+I_4),
\end{align}
where 
\begin{align}
\gamma_M^{-}=\lb(\frac{\gamma_{kd}}{1/\alpha_{ke}}\rb)^{-}=\max_{\substack{i\in \mathbb{S}\\i\ne k}}\lb\{\frac{\gamma_{id}}{1/\alpha_{ie}}\rb\}
=\max_{\substack{i\in \mathbb{S}\\i\ne k}}\lb\{\gamma_{id}\alpha_{ie}\rb\},
\end{align}
and $I_3$ and $I_4$ are expressed in (\ref{eq_tradimp2}) and (\ref{eq_tradimp2_1}) respectively.
% \begin{align}
% \label{eq_tradimp2} 
% I_3&=\int_{\rho-1}^\infty \int_{\frac{z-(\rho -1)}{\rho}}^\infty \int_0^{\alpha_{ke}\lambda} \int_{y/\alpha_{ke}}^\lambda 
% f_{\gamma_{kd}}(t)f_{\gamma_{M}^{-}}(y)f_{X}(x) \nn\\ 
% & \times f_{\gamma_{sd}}(z)  dt dy dx dz, \\
% I_4&= \int_{0}^{\rho-1} \int_{0}^\infty \int_0^{\alpha_{ke}\lambda} \int_{y/\alpha_{ke}}^\lambda 
% f_{\gamma_{kd}}(t)f_{\gamma_{M}^{-}}(y)f_{X}(x)\nn\\
% &\times f_{\gamma_{sd}}(z)  dt dy dx dz.
% \end{align}
The PDF of $\gamma_M^{-}$ can be easily obtained as in (\ref{eq_pdf}), with $\beta_m^{\prime}=\sum_{l=1}^m\beta_{i_ld}/\alpha_{le}$. 

$I_3$ and $I_4$ can be integrated to $I_3=-\sum\limits_{m=1}^{K-1} (-1)^m\mathbb {\sum}_m^{\prime} (P_{11}+P_{12}-P_{13})$ 
and $I_4=-\sum\limits_{m=1}^{K-1} (-1)^m\mathbb {\sum}_m^{\prime}(P_{21}+P_{22}-P_{23})$, respectively, where  $P_{11}, P_{12}, P_{13}, P_{21}, P_{22}, P_{23}$ 
are given in (\ref{eq_p1}) to (\ref{eq_p1_append}).

It is worth mentioning that (\ref{eq_p1}) to (\ref{eq_p1_append}) are valid for 
$(\alpha_{ke}\beta_m^{\prime}+\beta_{kd})\neq\beta_{sd}$ and $\beta_{kd}\neq\beta_{sd}$. When either ($\alpha_{ke}\beta_m^{\prime}+\beta_{kd})=\beta_{sd}$ or 
$\beta_{kd}=\beta_{sd}$, the analytical expressions can be similarly found after slight modifications.
\subsection{Optimal Selection (OS)}
\label{sec_optimal}
Optimal selection takes into account both main channel and $E$ channels quality. The relay is selected for which the secrecy rate is maximum, and  
\begin{align}
\label{eq_os1} &P_o^{\mathbb{S}}(R_s)=\mathbb{P}\lb[\max_{\forall k \in \mathbb{S}} \lb\{\frac{1+\gamma_{kd}+\gamma_{sd}}{1+\gamma_{ke}+\gamma_{se}}\rb\}\leq \rho\rb].
\end{align}
The above probability can be evaluated first for given $\gamma_{se}$ and $\gamma_{sd}$, and then by averaging over them. In this case, (\ref{eq_os1}) can be written 
as a product of individual probabilities
\begin{align}
\label{eq_os2} 
P_o^{\mathbb{S}}(R_s)&=\mathbb {E}_{\gamma_{se}}\mathbb {E}_{\gamma_{sd}} \lb[\prod_{\forall k \in \mathbb{S}}^K\mathbb{P}  \lb[\frac{1+\gamma_{kd}+\gamma_{sd}}
{1+\gamma_{ke}+\gamma_{se}}\leq \rho|\gamma_{se}, \gamma_{sd}\rb]\rb]\nn\\
&=\mathbb {E}_{\gamma_{se}}\mathbb {E}_{\gamma_{sd}} \lb[
\prod_{\forall k \in \mathbb{S}}^K\mathbb{P}\lb[\gamma_{kd}\leq \lb(\rho-1\rb)+\rho\lb(\gamma_{ke}+\gamma_{se}\rb)\rb.\rb.\nn\\
&\lb.\lb.-\gamma_{sd}|\gamma_{se}, \gamma_{sd}\rb]
\rb]= I_5+I_6,
% &=\prod_{\forall k \in \mathbb{S}}^K\lb[\int_{\frac{z-\rho y- (\rho-1)}{\rho}}^\infty \int_0^\lambda f_{\gamma_{kd}}(t)f_{\gamma_{ke}}(x) dt dx\rb]\nn\\
\end{align}
where  $I_5$ and $I_6$ are expressed in (\ref{eq_os3_1}) and (\ref{eq_os3_2}), respectively, with $\lambda= (\lb(\rho-1\rb)+\rho\lb(x+y\rb)-z)$, and $x$, $y$ and $z$ are realizations of the 
r.vs, $\gamma_{ke}$, $\gamma_{se}$, and $\gamma_{sd}$ respectively. The solution of $I_5$ and $I_6$ are provided in (\ref{eq_os4}) and  (\ref{eq_os5}), respectively, with 
$ {\sum}_m$ defined as \cite{Kundu_relsel}
\begin{align}
\mathbb {\sum}_m=\sum_{i_1=1 }^{K-(m-1)}
\sum_{i_2=i_1+1}^{K-(m-2)}\cdots \sum_{i_{m-1}=i_{m-2}+1}^{K-1} 
\sum_{i_m=i_{m-1}+1}^K.
\end{align}
We also define $\beta_m^{\prime}=\sum\limits_{l=1}^k \beta_{i_l d}$, $A_k^{\prime}=\prod\limits_{l=1}^k A_{i_l}$ and 
$A_k=\frac{\alpha_{ke}\exp{\lb(-\beta_{kd}(\rho-1)\rb)}}{\rho\beta_{kd}+\alpha_{ke}}$.
\section{Numerical and Simulation Results}
\label{sec_results}
This section describes numerical and simulation results. 
Unless otherwise mentioned, $1/\beta_{sd}=3$ dB, $1/\alpha_{se}=2$ dB, $\gamma_{th}=3$ dB, $N=4$, and $R=1$ bits per channel use (bpcu). 

\begin{figure}
\centering
\includegraphics[width=0.38\textwidth] {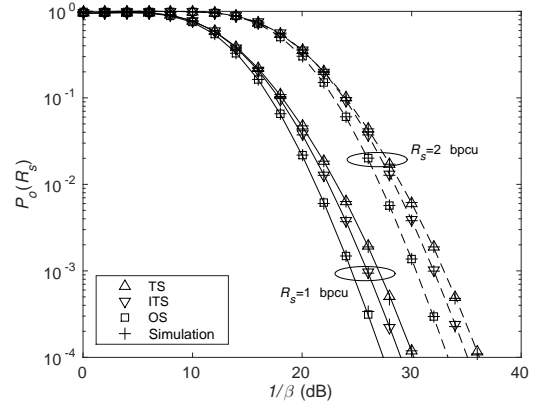}
\vspace*{0mm}
\caption{Secrecy outage probability versus average SNR for different $R_s$ values when links are non-identical. Solid lines are used for $R_s=1$ bpcu 
and dash lined for $R_s=2$ bpcu.}
\label{P14_FIG_1}
% \vspace*{-4mm}
\end{figure}

In Fig. \ref{P14_FIG_1}, the secrecy outage probabilities of the selection schemes TS, ITS and OS are plotted versus average SNR, $1/\beta$, 
for different rate requirements, $R_s=1,~ 2$ bpcu. Non-identical link parameters are considered, with 
$1/\beta_{sk}= 0.2/\beta,~0.6/\beta,~0.4/\beta,~0.8/\beta$, $1/\beta_{sk}=0.8/\beta,~ 0.4/\beta,~ 0.6/\beta,~ 0.2/\beta$, whereas $1/\alpha_{ke}=0,~ 3,~ 6,~ 9$ dB, respectively, 
for  $k=1,\cdots, 4$. As expected, OS works the best, followed by ITS, and TS is the worst. Additionally, it is 
worth noting that as $R_s$ increases, the secrecy outage probability deteriorates.

\begin{figure}
\centering	
\includegraphics[width=0.38\textwidth] {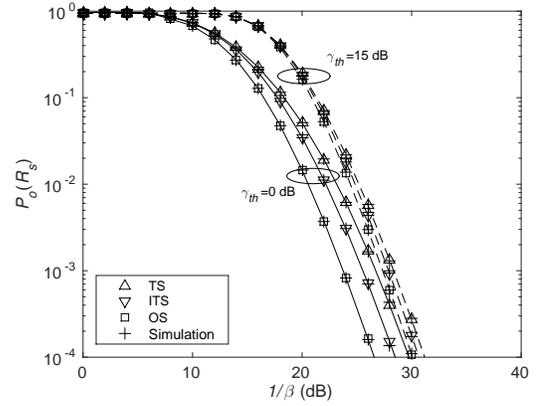}
\vspace*{0mm}
\caption{Secrecy outage probability versus average SNR for different SNR threshold values at the relays. Solid lines are used for $\gamma_{th}=0$ dB 
and dash lined for  $\gamma_{th}=15$ dB.}
\label{P14_FIG_3}
\vspace*{-5mm}
\end{figure}

Fig. \ref{P14_FIG_3} depicts the secrecy outage probabilities of the selection schemes TS, ITS and OS for two different $\gamma_{th}$ values, 
i.e., $\gamma_{th}=0, 15$  dB. It is assumed that the $S$-$R_k$ and $R_k$-$D$ link qualities are identical, i.e., 
$1/\beta_{sk}=1/\beta_{kd}=0.5/\beta$, while the $R_k$-$E$ link qualities are non-identical, i.e., 
$1/\alpha_{ke} = 0,~ 3,~ 6,~ 9$ dB for $k=1,\cdots, 4$. The secrecy outage probability deteriorates with the increase in $\gamma_{th}$. 
Higher values of $\gamma_{th}$ reduces the number of relays available for selection, hence the observation.

\begin{figure}
\centering
\includegraphics[width=0.38\textwidth] {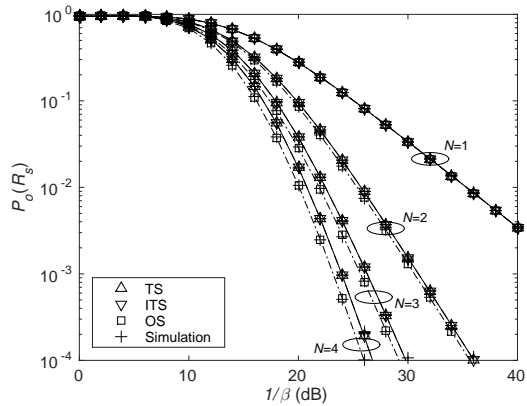}
\vspace*{0mm}
\caption{Secrecy outage probability versus average SNR for different $N$ values. Solid lines are used for TS, dash lines for ITS, 
and dash-dot lined for OS.}
\label{P14_FIG_2}
\vspace*{-0mm}
\end{figure}

In Fig. \ref{P14_FIG_2}, the secrecy outage probabilities of the selection schemes TS, ITS and OS are plotted versus the average SNR for increasing number of relays 
from $N=1$ to $N=4$. Identical link qualities are assumed as $1/\beta_{sk}=1/\beta_{kd}=0.5/\beta$ and $1/\alpha_{ke}=3$ dB for all $k$, respectively. 
It is clear that the performance improves as the number of relays increase. 
The slope of the curves also increases with increasing the number of relays, which means that the diversity order of the secrecy outage probability improves with $N$.
An important observation is that the improvement obtained by increasing the number of relays follows the laws of diminishing return. 
Furthermore, as the $R_k$-$E$ link qualities are identical for all $k$, the ITS selection scheme can not provide better performance than the TS selection scheme and 
merges with TS. It is worth noting that the performances of the TS and ITS schemes do not merge in Fig. \ref{P14_FIG_1} and Fig. \ref{P14_FIG_3}, respectively. 

\begin{figure}
\centering
\includegraphics[width=0.38\textwidth] {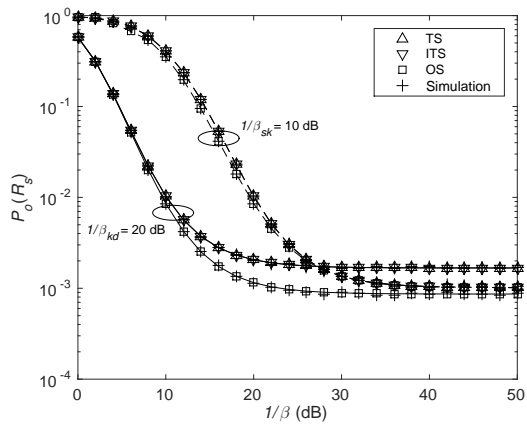}
\vspace*{0mm}
\caption{Secrecy outage probability versus average SNR for unbalanced $S$-$R_k$ and $R_k$-$D$ links. Solid lines are used for $1/\beta_{kd}=20$ dB 
and dash lined for  $1/\beta_{sk}=10$ dB.}
\label{P14_FIG_4}
\vspace*{-6mm}
\end{figure}

Fig. \ref{P14_FIG_4} shows the secrecy outage probabilities of the selection schemes TS, ITS and OS when the links $S$-$R_k$ and $R_k$-$D$ are unbalanced. 
Two cases are considered, as follows: Case 1, when $1/\beta_{kd}=20$ dB and $1/\beta_{sk}=1/\beta$, and Case 2, 
when $1/\beta_{sk}=10$ dB and $1/\beta_{kd}=1/\beta$, for $k=1,\cdots, 4$. The results are obtained assuming identical eavesdropper link qualities, $1/\alpha_{ke}=3$ dB, for all $k$. 
It is observed that the secrecy outage probability saturates to a particular value depending on the values of 
$1/\beta_{sk}$ or $1/\beta_{kd}$. This indicates that either link $S$-$R_k$ or $R_k$-$D$, for all $k$, can limit the 
the secrecy outage probability. 
Furthermore, it can be observed that when $1/\beta_{sk}=10$ dB, the performance of the relay selection 
schemes saturates to the same value, while the performance saturates to different values when $1/\beta_{kd}=20$ dB. 
When the $S$-$R_k$ link quality improves, the number of relays that exceed $\gamma_{th}$ increases. As the relay selection schemes can take increased advantage 
when there are more relays to choose from, performance saturates to different values in Case 1 depending on the selection scheme. 

It should be noted that, in all figures, simulation results are in agreement with numerical results. This validates our analysis in Section \ref{sec_multrel}.

\section{Conclusion}
\label{sec_conclusions}
Three relay selection schemes, namely traditional, improved traditional, and optimal are proposed to enhance the secrecy outage probability 
using threshold-selection DF relays. The secrecy outage probability is derived in closed-form assuming the most practical scenario of independent 
but non-identical fading channels and including direct links from the source to destination and eavesdropper. 
It is found that by increasing the number of relays, the diversity gain of the secrecy outage probability 
can be increased. On the other hand, higher SNR threshold at the relays can decrease the secrecy performance. It is observed that the improved 
traditional relay selection can outperform the traditional relay selection only if the eavesdropper links are non-identical. It is also noticed 
that the secrecy outage probability is limited by either the source to relay or the relay to destination link quality.
\begin{table*}
\begin{tabular}{m{\textwidth}}
% {p{\textwidth}}
{
\begin{align}
\label{eq_td_i1}
I_1&=-\sum\limits_{m=1}^{N-1}\lb(-1\rb)^{m}\mathbb {\sum}_m^{\prime}
\frac{B_1 \beta_{sd}\exp(-\beta_{sd}(\rho-1))}{\alpha_{se}/\rho+\beta_{sd}}
\lb[ 
\frac{\beta_m^{\prime}}{\alpha_{se}\lb(\beta_m^{\prime}+\beta_{kd}\rb)}
+\frac{\beta_{kd}}{\lb(\beta_m^{\prime}+\beta_{kd}\rb)\lb(\rho\lb(\beta_m^{\prime}+\beta_{kd}\rb)+\alpha_{se} \rb)}
-\frac{1}{\rho\beta_{kd}+\alpha_{se}}
\rb] \nn\\
&-\sum\limits_{m=1}^{N-1}\lb(-1\rb)^{m}\mathbb {\sum}_m^{\prime}
\frac{B_2 \beta_{sd}\exp(-\beta_{sd}(\rho-1))}{\alpha_{ke}/\rho+\beta_{sd}}
\lb[ 
\frac{\beta_m^{\prime}}{\alpha_{ke}\lb(\beta_m^{\prime}+\beta_{kd}\rb)}
+\frac{\beta_{kd}}{\lb(\beta_m^{\prime}+\beta_{kd}\rb)\lb(\rho\lb(\beta_m^{\prime}+\beta_{kd}\rb)+\alpha_{ke} \rb)}
-\frac{1}{\rho\beta_{kd}+\alpha_{ke}}
\rb].\\
% \end{align}
% }
% {
% \begin{align}
\label{eq_td_i2}
I_2&=-\sum\limits_{m=1}^{N-1}\lb(-1\rb)^{m}\mathbb {\sum}_m^{\prime}
B_1 \lb[ 
\frac{\beta_m^{\prime}\lb(1-\exp\lb(-\beta_{sd}(\rho-1)\rb)\rb)}{\alpha_{se}\lb(\beta_m^{\prime}+\beta_{kd}\rb)}
-\frac{\beta_{sd}\exp(-\beta_{kd}(R_s-1))\lb(\exp((\beta_{kd}-\beta_{sd})(R_s-1))-1\rb)}{\lb(\rho\beta_{kd}+\alpha_{se}\rb)\lb(\beta_{kd}-\beta_{sd} \rb)}
\rb.\nn\\
&\lb.+\frac{\beta_{sd}\beta_{kd}\exp(-(\beta_{kd}+\beta_m^{\prime})(\rho-1))
\lb(\exp((\beta_{kd}+\beta_m^{\prime}-\beta_{sd})(\rho-1))-1\rb)}{\lb(\beta_m^{\prime}+\beta_{kd}\rb)\lb(\rho\lb(\beta_m^{\prime}
+\beta_{kd}\rb)+\alpha_{se} \rb)\lb(\beta_m^{\prime}+\beta_{kd}-\beta_{sd}\rb)}
\rb] \nn\\
&-\sum\limits_{m=1}^{N-1}\lb(-1\rb)^{m}\mathbb {\sum}_m^{\prime}
B_2\lb[ 
\frac{\beta_m^{\prime}\lb(1-\exp\lb(-\beta_{sd}(\rho-1)\rb)\rb)}{\alpha_{ke}\lb(\beta_m^{\prime}+\beta_{kd}\rb)}
-\frac{\beta_{sd}\exp(-\beta_{kd}(R_s-1))\lb(\exp((\beta_{kd}-\beta_{sd})(R_s-1))-1\rb)}{\lb(\rho\beta_{kd}+\alpha_{ke}\rb)\lb(\beta_{kd}-\beta_{sd} \rb)}
\rb.\nn\\
&\lb.+\frac{\beta_{sd}\beta_{kd}\exp(-(\beta_{kd}+\beta_m^{\prime})(\rho-1))
\lb(\exp((\beta_{kd}+\beta_m^{\prime}-\beta_{sd})(\rho-1))-1\rb)}{\lb(\beta_m^{\prime}+\beta_{kd}\rb)\lb(\rho\lb(\beta_m^{\prime}
+\beta_{kd}\rb)+\alpha_{ke} \rb)\lb(\beta_m^{\prime}+\beta_{kd}-\beta_{sd}\rb)}
\rb].\\\nn\\
% \end{align}
% }
% \\
\hline \nn\\
% \hrule
% {
% \begin{align}
\label{eq_tradimp2} 
I_3&=\int_{\rho-1}^\infty \int_{\frac{z-(\rho -1)}{\rho}}^\infty \int_0^{\alpha_{ke}\lambda} \int_{y/\alpha_{ke}}^\lambda 
f_{\gamma_{kd}}(t)f_{\gamma_{M}^{-}}(y)f_{X}(x)f_{\gamma_{sd}}(z)  dt dy dx dz. \\
\label{eq_tradimp2_1} 
I_4&= \int_{0}^{\rho-1} \int_{0}^\infty \int_0^{\alpha_{ke}\lambda} \int_{y/\alpha_{ke}}^\lambda 
f_{\gamma_{kd}}(t)f_{\gamma_{M}^{-}}(y)f_{X}(x)f_{\gamma_{sd}}(z)  dt dy dx dz.
% \end{align}
% }
% \\\hline
% {
% \begin{align}
\\\nn\\
\hline 
\nn\\
\label{eq_p1}
P_{11}&=\frac{\beta_{sd}\alpha_{ke}\beta_m^{\prime}\exp(-\beta_{sd}(\rho-1))}{\alpha_{ke}\beta_m^{\prime}+\beta_{kd}}
\lb[\frac{B_1}{\alpha_{se}\lb(\alpha_{se}/\rho+\beta_{sd}\rb)}
+\frac{B_2}{\alpha_{ke}\lb(\alpha_{ke}/\rho+\beta_{sd}\rb)}\rb]. \\
P_{12}&=\frac{\beta_{sd}\beta_{kd}\exp(-\beta_{sd}(\rho-1))}{\alpha_{ke}\beta_m^{\prime}+\beta_{kd}}
\lb[\frac{B_1}{\lb(\alpha_{se}/\rho+\beta_{sd}\rb)\lb(\rho(\alpha_{ke}\beta_m^{\prime}+\beta_{kd})+\alpha_{se}\rb)}
+\frac{B_2}{\lb(\alpha_{ke}/\rho+\beta_{sd}\rb)\lb(\rho(\alpha_{ke}\beta_m^{\prime}+\beta_{kd})+\alpha_{ke}\rb)}\rb]. \\
P_{13}&=\beta_{sd}\exp(-\beta_{sd}(\rho-1))
\lb[\frac{B_1}{\lb(\alpha_{se}/\rho+\beta_{sd}\rb)\lb(\rho\beta_{kd}+\alpha_{se}\rb)}
+\frac{B_2}{\lb(\alpha_{ke}/\rho+\beta_{sd}\rb)\lb(\rho\beta_{kd}+\alpha_{ke}\rb)}\rb]. \\
P_{21}&=\frac{\alpha_{ke}\beta_m^{\prime}}{\alpha_{ke}\beta_m^{\prime}+\beta_{kd}}\lb(1-\exp(-\beta_{sd}(\rho-1))\rb).\\
P_{22}&=\frac{\beta_{sd}\beta_{kd}\exp(-(\alpha_{ke}\beta_m^{\prime}+\beta_{kd})(\rho-1))}{\alpha_{ke}\beta_m^{\prime}+\beta_{kd}}
\lb[\frac{B_1}{\lb(\rho(\alpha_{ke}\beta_m^{\prime}+\beta_{kd})+\alpha_{se}\rb)}
+\frac{B_2}{\lb(\rho(\alpha_{ke}\beta_m^{\prime}+\beta_{kd})+\alpha_{ke}\rb)}\rb]\nn\\
&\times\frac{\exp(-(\alpha_{ke}\beta_m^{\prime}+\beta_{kd}-\beta_{sd})(\rho-1))-1}{(\alpha_{ke}\beta_m^{\prime}+\beta_{kd}-\beta_{sd})}. \\
\label{eq_p1_append}
P_{23}&=\beta_{sd}\exp(-\beta_{kd}(\rho-1))
\lb[\frac{B_1}{\lb(\rho\beta_{kd}+\alpha_{se}\rb)}
+\frac{B_2}{\lb(\rho\beta_{kd}+\alpha_{ke}\rb)}\rb]\frac{\exp\lb(\lb(\beta_{kd}-\beta_{sd}\rb)(\rho-1)\rb)-1}{\lb(\beta_{kd}-\beta_{sd}\rb)}. 
% \end{align}
% }
% \\\hline
% {
% \begin{align}
% \label{eq_pdfmin2}
% I_1 &=\frac{\alpha_{se}\alpha_{ke}}{\alpha+\alpha_{ke}}\lb[\frac{B_1\exp\lb(-\beta_{sd}(\rho-1)\rb)}{\lb(\rho\beta_{sd}+\alpha_{se}\rb)\lb((\alpha_{ke}+\alpha)/\rho+\beta_{sd}\rb)}
% +\frac{B_2\exp\lb(-\beta_{kd}(\rho-1)\rb)}{\lb(\rho\beta_{kd}+\alpha_{se}\rb)\lb((\alpha_{ke}+\alpha)/\rho+\beta_{kd}\rb)}  \rb], \nn\\
% I_2 &=\frac{\alpha_{ke}}{\alpha+\alpha_{ke}}\lb[\frac{B_1}{\beta_{sd}}+\frac{B_2}{\beta_{kd}}
% -\frac{B_1\alpha_{se}\exp\lb(-\beta_{sd}(\rho-1)\rb)}{\lb(\rho\beta_{sd}+\alpha_{se}\rb)\beta_{sd}}
% -\frac{B_2\alpha_{se}\exp\lb(-\beta_{kd}(\rho-1)\rb)}{\lb(\rho\beta_{kd}+\alpha_{se}\rb)\beta_{kd}}  \rb].
% \end{align}
% }\\\hline
% {
% \begin{align}
\\\nn\\
\hline 
\nn\\
\label{eq_os3_1} 
I_5&= \int_0^\infty \int_{\rho y+ (\rho-1)}^\infty   \lb[\prod_{\forall k \in \mathbb{S}}^K\lb(\int_{\frac{z-\rho y- (\rho-1)}{\rho}}^\infty \int_0^\lambda 
f_{\gamma_{kd}}(t)f_{\gamma_{ke}}(x) dt dx\rb)\rb] f_{\gamma_{sd}}(z)f_{\gamma_{se}}(y)dz dy \\
\label{eq_os4}
&=\frac{\beta_{sd}\alpha_{se}}{\lb(\sum_{k=1}^N \alpha_{ke}/\rho +\beta_{sd}\rb)\lb(\rho  \beta_{sd}+\alpha_{se}\rb)} \prod_{k=1}^N \frac{\rho \beta_{kd}}{\rho  \beta_{kd}+ \alpha_{ke}}.\\
\label{eq_os3_2}
I_6&= \int_0^\infty \int_{0}^{\rho y+ (\rho-1)}  \lb[\prod_{\forall k \in \mathbb{S}}^K\lb(\int_0^\infty \int_0^\lambda 
f_{\gamma_{kd}}(t)f_{\gamma_{ke}}(x) dt dx\rb)\rb] f_{\gamma_{sd}}(z)f_{\gamma_{se}}(y)dz dy \\
\label{eq_os5}
&=1-\frac{\alpha_{se}\exp(-\beta_{sd}(\rho-1))}{\rho\beta_{sd}+\alpha_{se}}
+\sum\limits_{m=1}^{N}\lb(-1\rb)^{m}\mathbb {\sum}_m  
\lb[\frac{A_m^{\prime}\beta_{sd}\alpha_{se}\exp\lb((\beta_m^{\prime}-\beta_{sd})(\rho-1)\rb)}{(\beta_m^{\prime}-\beta_{sd})(\rho\beta_{sd}+\alpha_{se})}
- \frac{A_m^{\prime}\beta_{sd}\alpha_{se}}{(\beta_m^{\prime}-\beta_{sd})(\rho\beta_m^{\prime}+\alpha_{se})}\rb].
\end{align}
}
% \\\hline
% {\begin{align}
% }
\end{tabular}
\end{table*}
% \end{small}

\bibliographystyle{IEEEtran}
\bibliography{IEEEabrv,MYALL_REFERENCE}
\end{document}